\documentclass[preprint,showpacs,preprintnumbers,amsmath,amssymb,a4paper,superscriptaddress]{revtex4}

\usepackage{graphicx}
\usepackage{dcolumn}
\usepackage{bm}

\begin{document}

\title{Suppression of Quantum Scattering in Strongly Confined Systems}

\author{J.I. Kim}
\affiliation{Conf. Altanova, Dept. de Pesquisas,
R. Silva Teles 712, CEP 03026-000, Br\'{a}s, S\~{a}o Paulo, SP, Brasil,}
\email{ji_il_kim@yahoo.com}
\author{V.S. Melezhik}
\affiliation{Bogoliubov Laboratory of Theoretical Physics, Joint Institute for Nuclear Research,
Dubna, Moscow Region 141980, Russian Federation,}
\email{melezhik@thsun1.jinr.ru}
\author{P. Schmelcher}
\affiliation{Physikalisches Institut, Universit\"{a}t Heidelberg, Philosophenweg 12, 69120
Heidelberg, Germany,}
\affiliation{Theoretische Chemie, Institut f\"{u}r Physikalische Chemie, Universit\"{a}t
Heidelberg, Im Neuenheimer Feld 229, 69120 Heidelberg, Germany.}
\email{peter.schmelcher@tc.pci.uni-heidelberg.de}

\date{7 April 2006}

\begin{abstract}
We demonstrate that scattering of particles strongly interacting 
in three dimensions (3D) can be suppressed at low energies in a 
quasi-one dimensional (1D) confinement. The underlying mechanism is the 
interference of the $s$- and $p$-wave scattering contributions
with large $s$- and $p$-wave 3D scattering lengths being a necessary
prerequisite. This low-dimensional quantum scattering effect might
be useful in ``interacting" quasi-1D ultracold atomic gases, guided atom
interferometry and impurity scattering in strongly confined quantum
wire-based electronic devices.
\end{abstract}

\pacs{03.75.Be, 34.10.+x, 34.50.-s}

\maketitle

In a scattering process the boundary conditions play a key role and can substantially
affect the scattering outcome.
In case of asymptotically free motion and
for spherically symmetric potentials, it is natural to analyze the
scattering processes using partial waves belonging to certain angular momenta. Consider
then a scattering event in 3D between two distinguishable particles with relative momentum
$k$. Adding contributions of higher partial waves with increasing orbital angular momentum $l$ and
phase shifts $\delta_l$ can only {\em increase} the total cross-section, which in 3D is given by
\begin{equation}
\label{sigma}
\sigma=\frac{4\pi}{k^2}\sum_l (2l+1)\sin^2{\delta_l} .
\end{equation}
A pure $s$-wave ($l=0$) approximation may not be accurate even at low energies for potentials that
possess a significant spectral structure,
e.g., impurity or dopant scattering in some bulk
solid state systems~\cite{conwell1950a,ridley2000a} or $p$-wave ($l = 1$) interactions in ultracold
atomic physics
\cite{Santos00,Santos03,Stuhler05,ticknor2004a,stock2005a,pricoupenko2005b,idziaszek2006a,guenter2005a}.
Such additional scattering terms for different partial waves could lead to increased
heat dissipation along bulk conduction channels or to higher phase degradation in atom
interferometry \cite{gir2002,jas2004,schum} due to atomic collisions.

The scattering process can change dramatically
if it takes place in a partially confined geometry , such as quasi-1D systems.
The latter situation is encountered in certain quantum
wires~\cite{bryant1984a,chu1989a,chuu1992a,cui2003a} for electronic transport or in wave
guides for ultracold
atoms~\cite{olshanii1998a,bergeman2003,granger2004a,folman2002a,guenter2005a}. Of particular
interest is the single mode regime, where only the ground state
(hereafter indicated by the index 0)
with respect to the confining directions can be actually populated. Then,
instead of $\sigma$, the quasi-1D total cross-section is the reflection coefficient $R$, which can be written
in terms of the effective quasi-1D forward ($z\rightarrow +\infty$) scattering amplitude $f_0^+$
~\cite{olshanii1998a,bergeman2003,granger2004a,kim2005a},
\begin{equation}
\label{t}
R= 1 - |1+f_0^+|^2,
\hspace{3em}
f_0^+=f_{0g}+f_{0u},
\end{equation}
where $f_{0g}$ and $f_{0u}$ are even and odd quasi-1D amplitudes
containing even and odd partial $l$-waves, respectively. In a
seminal work~\cite{olshanii1998a}, in the $s$-wave approximation
for which $f_{0u} \approx 0$, a resonance $R \approx 1$ has been
predicted with $f_{0g} \approx -1$. The corresponding low energy
transport is thus almost blocked and the longitudinal motion
exhibits a near infinite effective quasi-1D scattering: the
original confinement induced resonance (CIR)~\cite{olshanii1998a}.
An account of the center of mass motion of colliding particles in
the pure $s$-wave zero-range approximation is provided
in~\cite{peano2005a}. On the other hand, higher partial wave
contributions like the next $p$-wave
\cite{ticknor2004a,stock2005a,pricoupenko2005b,idziaszek2006a,guenter2005a,granger2004a,kim2005a}
in $f_{0u}$ can also exhibit a resonance~\cite{granger2004a} with
$f_{0u} \approx -1$. This is another diverging effective
interaction, but in the odd sector~\cite{granger2004a}.

The present work addresses the novel situation where a {\em
simultaneous} resonance in the $s$- and $p$-wave scattering occurs
such that $R\approx 0$. We show that the even and odd
contributions to the scattering amplitude interfere and may
provide effectively a quasi-1D gas of non-interacting atoms or a
free flow of carriers in a quantum wire. Note, that this free flow
occurs in spite of a strong interaction in 3D. We remark that for
cold atoms dc electric field can supply us with hybridized
Feshbach resonances providing a situation where both $s$- and
$p$-wave scattering becomes simultaneously
resonant~\cite{krems2006a}. The above-mentioned complementary or
``dual" CIR due to both large $s$- and $p$-wave scattering is
established by extracting the transmission $T= 1-R =
|1+f_{0}^+|^2$ from the wave packet dynamical solution of the
time-dependent Schr\"{o}dinger equation for two particles with
masses $m_1\neq m_2$ and by highlighting the key concepts at play.
The external transverse confinement is assumed to be parabolic
$U_i(\rho_i)=\frac{1}{2}m_i\omega_i^2\rho_i^2$ where
$\bm{r}_i=(\bm{\rho}_i,z_i)$, $i=1,2$, are the coordinates of the
particles. Their interaction is the screened Coulomb potential
($V_0<0$)
\begin{equation}
\label{yukawa}
V(r)=V_0\frac{r_0}{r}\, e^{-r/r_0},
\end{equation}
where $r=|\bm{r}_1-\bm{r}_2|$ and $r_0$ is the screening length.
Note that only for $\omega_1=\omega_2$ the center of mass
decouples and e.g. an impurity-like scattering can be described
solely by the relative coordinates. If $\omega_1\neq\omega_2$, a
two-species gas of ultracold trapped atoms, e.g., $^{40}$K,
$^{85}$Rb, $^{87}$Rb and combinations thereof, can be
described qualitatively, since only the
low-energy asymptotic behaviour of the scattering process is
required, which should depend mostly on properties such as the
scattering lengths and the range of the potential $V$.
Independent simulations with other potentials, e.g.,
$C_6$-$C_{12}$~\cite{bergeman2003} or hyperbolic-cosine~\cite{granger2004a}
are expected to provide the same behaviour and properties.
Additionally Eq.(\ref{yukawa}) allows for a
good control of the low energy scattering lengths by tuning its
spectrum via $V_0$.
The computation of
$T$ is first performed for $\omega_1=\omega_2$
and second for
$\omega_1\neq\omega_2$, including the coupling to the center of
mass.

In the Schr\"{o}dinger equation, the center of mass coordinates
$\bm{R}=\frac{m_1}{M}\bm{r}_1+\frac{m_2}{M}\bm{r}_2$, with
$M=m_1+m_2$, are employed in the cylindrical representation
$(\rho_R,\phi_R,Z)$ while for the relative coordinates
$\bm{r}=\bm{r}_1-\bm{r}_2$, depending on the representation
both cylindrical $(\rho,\phi,z)$ and
spherical coordinates $(r,\theta,\phi)$ are needed. By symmetry,
the motion with respect to the degree of freedom $Z$ can be
decoupled from the problem and will be omitted. In addition, due
to the conservation of the $z$-component of the total angular
momentum, namely, $L_{1z}+L_{2z}=-i(\partial_{\phi_1}+
\partial_{\phi_2})=-i(\partial_{\phi} + \partial_{\phi_R})$,
$\phi_i$ being the azimuthal angle of $\bm{r}_i$ ($\hbar=1$), the number of
independent variables can be further reduced in a frame
co-rotating with the center of mass around the symmetry
$z$-axis. This is accomplished by the unitary
transformation~\cite{bock2005a}
\begin{equation}
\label{rotation}
\mathcal{\hat{U}}=e^{i\phi_R L_z},
\end{equation}
where $L_z=-i\partial_{\phi}$. Expressing the total Hamiltonian in terms of
$\rho_R, \phi_R, r, \rho, \theta$ and $\phi$, the net effect of this transformation is to shift
$\partial_{\phi_R}$ to $\partial_{\phi_R} - \partial_{\phi}$
in the kinetic operator for the center of mass and $\phi$ to $\phi+\phi_R$ in the potential operator
that couples $\rho$ to $\rho_R$.
The rotated Hamiltonian without $V$ becomes then
\begin{equation}
\label{h0}
H_0 = H_M + H_\mu + W\, .
\end{equation}
Here $H_M=-(\partial^2_{\rho_R} + 1/4\rho_R^2)/2M -
(\partial_{\phi_R}-\partial_\phi)^2/2M\rho_R^2$,
$W=\frac{1}{2}M\omega_M^2\rho_R^2 +
\frac{1}{2}\mu\omega_\mu^2\rho^2 +
\mu(\omega_1^2-\omega_2^2)\rho\rho_R\cos{\phi}$ and
$H_\mu=-\partial^2_r/2\mu+L^2/2\mu r^ 2$, where $L^2$ is the
square of the orbital angular momentum of the relative coordinates
with eigenvalues $l(l+1)$, $\mu=m_1m_2/M$ is the reduced mass,
$\omega_{\mu}^2=(m_2/M)\omega_1^2+(m_1/M)\omega_2^2$  and
$\omega_M^2=(m_1/M)\omega_1^2+(m_2/M)\omega_2^2$ characterize the
confinement oscillator frequencies for the relative and center of
mass coordinates, respectively, whereas $V$ is invariant with
respect to the transformation $\mathcal{\hat{U}}$.
The derivative $\partial_\phi$ in the second term of $H_M$ is a Coriolis coupling term, which is
characteristic for the rotating frame of reference and appears as an additional kinetic energy
operator.

The initial wave packet at $t=0$ possesses a Gaussian shape
of width $a_z$ and momentum $k_0\equiv \sqrt{2\mu\,\varepsilon}>0$ for the unconfined $z$-motion
\begin{equation}
\label{t0}
\Psi(0)=N r\sqrt{\rho_R}\ [\mathcal{\hat{U}}\Phi]
\ e^{-(z-z_0)^2/2a_z^2}\ e^{izk_0},
\end{equation}
$N$ is the overall normalization constant
defined as $\langle\Psi(0)|\Psi(0)\rangle = \int_0^\infty dr\int_0^\infty d\rho_R \int_0^\pi
sin\theta d\theta \int_0^{2\pi} d\phi \Psi^*(0)\Psi(0) = 1$
and the particles are
separated by $z_0<0$ while their confined motions $\bm{\rho_i}$ are
in the respective harmonic oscillator ground state
$\Phi=\Phi(\rho_1,\rho_2)=e^{-(\rho_1^2/a_1^2 +
\rho_2^2/a_2^2)/2}$ with $a_i=(1/m_i\omega_i)^{1/2}$.
By expressing $\rho_i$ in terms of the variables $\rho_R,
\phi_R, \rho$ and $\phi$, the action of $\mathcal{\hat{U}}$ is to
simply shift $\phi$ to $\phi+\phi_R$ in the exponent of $\Phi$, as
can be seen by using the basis of eigenstates of $L_z$.  As a
result, the $\phi_R$ variable drops from the initial state
$\Psi(0)$ and the Schr\"{o}dinger equation
$i\dot{\Psi}(t) = \left[ H_0 + V \right]\Psi(t)$
reduces to the subspace of the four variables $(\rho_R, \bm{r})$.
The $z-$component of the total angular momentum is in the co-rotating frame
given by $L_Z=-i\partial_{\phi_R}$. According to our initial wave packet in Eq.(\ref{t0})
we assume now a vanishing total angular momentum $L_Z \Psi = 0$ and therefore omit the corresponding
derivative terms in $H_M$.

The extraction of $T$ from $\Psi(t)$ follows from the asymptotics at large times $t$, when $\Psi(0)$ has
been split by $V$ into a backward and a forward scattered part outside the range of $V$
ideally for $z\rightarrow\mp\infty$, respectively. It can be obtained from the forward scattered part by
integrating the asymptotic density $|\Psi(t)|^2$ over the half-space $z\geq r_0$ or from the projection
$|\langle\Psi_0(t)|\Psi(t)\rangle|^2
\mathop{\longrightarrow}\limits_{t \rightarrow
\infty}|1+f_0^+|^2$
onto the {\em unscattered} large $t$ evolution of $\Psi(0)$
under $H_0$ alone. In the limit $a_z \rightarrow \infty$ our wave packet scattering corresponds
asymptotically ($t \rightarrow \infty$) to the monoenergetic stationary scattering situation thereby
providing the energy or momentum resolved scattering parameter $T=T(k)$.

The present computational scheme to solve for $\Psi(t)$ is a four
dimensional extension of a three dimensional method originally
developed to treat, among others, bound-bound and bound-continuum
transitions for atomic systems in external fields
\cite{melezhik1997a,melezhik1999b,melezhik2003a}. In this scheme,
an angular basis $f_j(\theta,\phi)$ is
constructed on the grid $(\theta_j,\phi_j)$ using the exponentials
$e^{im\phi}$ and the Legendre function $P_l^m(\theta)$, such that
the only non-diagonal terms are the angular parts of the kinetic
energy operators, namely, $L^2/2\mu r^ 2$ and  $-
\partial^2_\phi/2M\rho_ R^2$. These, in turn, can be diagonalized
efficiently if one employs additionally a simple unitary
transformation between the two basis $f_j$ and $e^{im\phi}$,
$P_l^m(\theta)$~\cite{melezhik1997a,melezhik2003a}. Our approach
is reminiscent of a two dimensional discrete variable
representation (DVR) \cite{lill2000a} and the values of the wave
function $\Psi(t)$ are given on an angular grid
$(\theta_j,\phi_j)$ whose total number of grid points equals the
number of basis functions $f_j$. The solution is then propagated
in time using a component-by-component split-operator
method~\cite{melezhik1997a,melezhik2003a} according to
$\Psi(t+\Delta t)\approx e^{-\frac{i}{2}W\Delta
t}e^{-i(H_\mu+V)\Delta t}e^{-iH_M\Delta t} e^{-\frac{i}{2}W\Delta
t}\Psi(t)$.

To be specific the parameters are chosen as follows. $r_0=1$ defines the length scale,
$m_1/m_2=40/87$ is the mass ratio (e.g., $^{40}$K and $^{87}$Rb)
and the reduced mass is taken as the unit mass $\mu=1$.
We define $\omega=(\omega_1+\omega_2)/2$ and the confinement length
$a_\perp\equiv\sqrt{1/\mu\omega}$. For the case of a decoupled center of mass motion,
$\omega_1=\omega_2=\omega$, whereas for the coupled case,
$\omega_1=1.35 \omega_2$. The low longitudinal energy
$\varepsilon=0.002$ guarantees the single mode regime $ \omega <
E=\omega+\varepsilon <3\omega$ for the region of $0.002\leq\omega \leq 0.02$ we consider.

For $-V_0 < 9$ (in units of  $\hbar^2/\mu r_0^2$), there exists no bound-state of $V(r)+l(l+1)/2\mu
r^2$ for $l\geq 2$. Thus only the $s$- and $p$-wave 3D scattering lengths $a_s$ and $a_p$ can be
large. These are computed in the limit $k \rightarrow 0$ by $k \cot{\delta_0(k)} = -1/a_s$ and $k^3
\cot{\delta_1(k)} = -1/V_p$ , with $V_p \equiv a_p^3$.

\begin{figure}[t]
\includegraphics[scale=0.79]{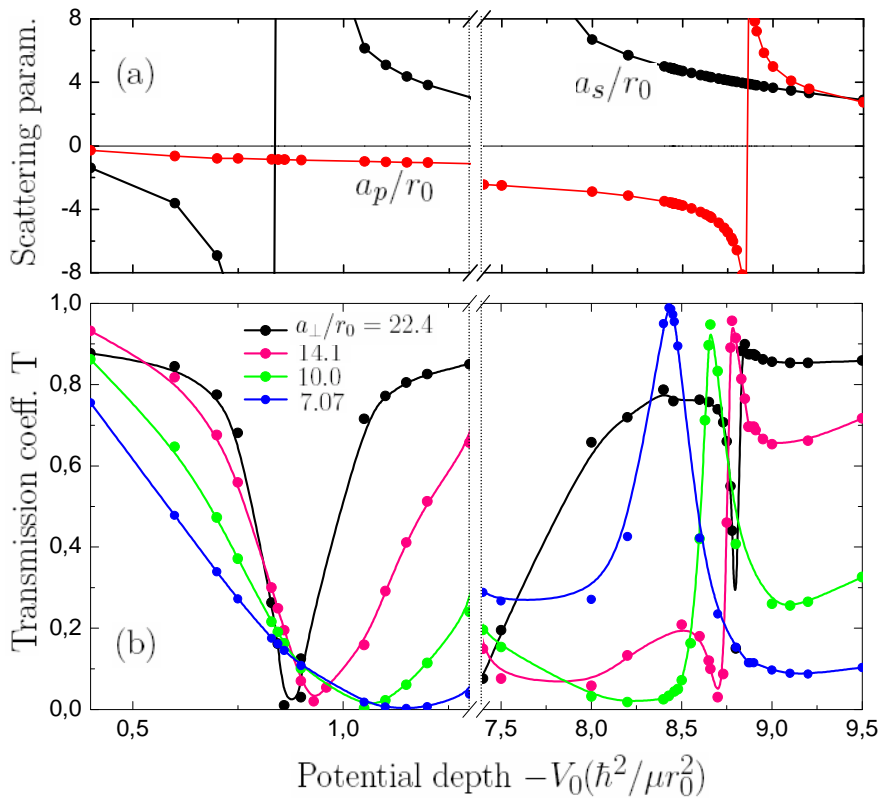}
\caption{ \label{fig2} (color online)
({\bf a}) The 3D scattering lengths $a_s$ (black) and $a_p$ (red)
and ({\bf b}) the quasi-1D transmission coefficients $T$ for
a few values of $a_{\perp}$ and fixed $\varepsilon=0.002 (k_0=0.0632)$. Discussion
see text.}
\end{figure}

First let us consider the confined scattering process
for the case of a decoupled center of mass motion $\omega_1=\omega_2=\omega$. The
results for $T$ are shown in Fig.~\ref{fig2} together with the behaviour
of the
3D
scattering lengths $a_s$ and $a_p$.
These
are tuned by changing $V_0$ in two regions where
correspondingly only $a_s$ ($-V_0\sim 1$) or both $a_s$ and $a_p$
($-V_0\sim 7.5-9.5$) are large
, i.e., on
the order of $a_\perp$.
In the first region we see the
well-known $s$-wave CIR which leads to a minimum
$T\approx 0$ when the ratio $a_\perp/a_s$ approaches
$a_\perp/a_s \approx 1.46$ in agreement with
\cite{olshanii1998a,bergeman2003}. In the second region, however, remarkable peaks
$T\approx 1$ in the transmission are observed being most pronounced for small momenta
$k_0$ (small
$a_\perp$).
Considering in this case pure $s$-wave scattering would yield $T\approx 0$~\cite{olshanii1998a}
in contrast to the behaviour $T\approx 1$ shown in Fig.~\ref{fig2}.
Clearly this is due to the fact that both $a_s$ and $a_p$ become large and contribute
equally to the scattering process. We therefore encounter the
peculiar situation of an almost complete transmission, i.e. free flow, in spite of the
strong interaction between the scattering partners
in free space.

\begin{figure}[t]
\includegraphics[scale=0.70]{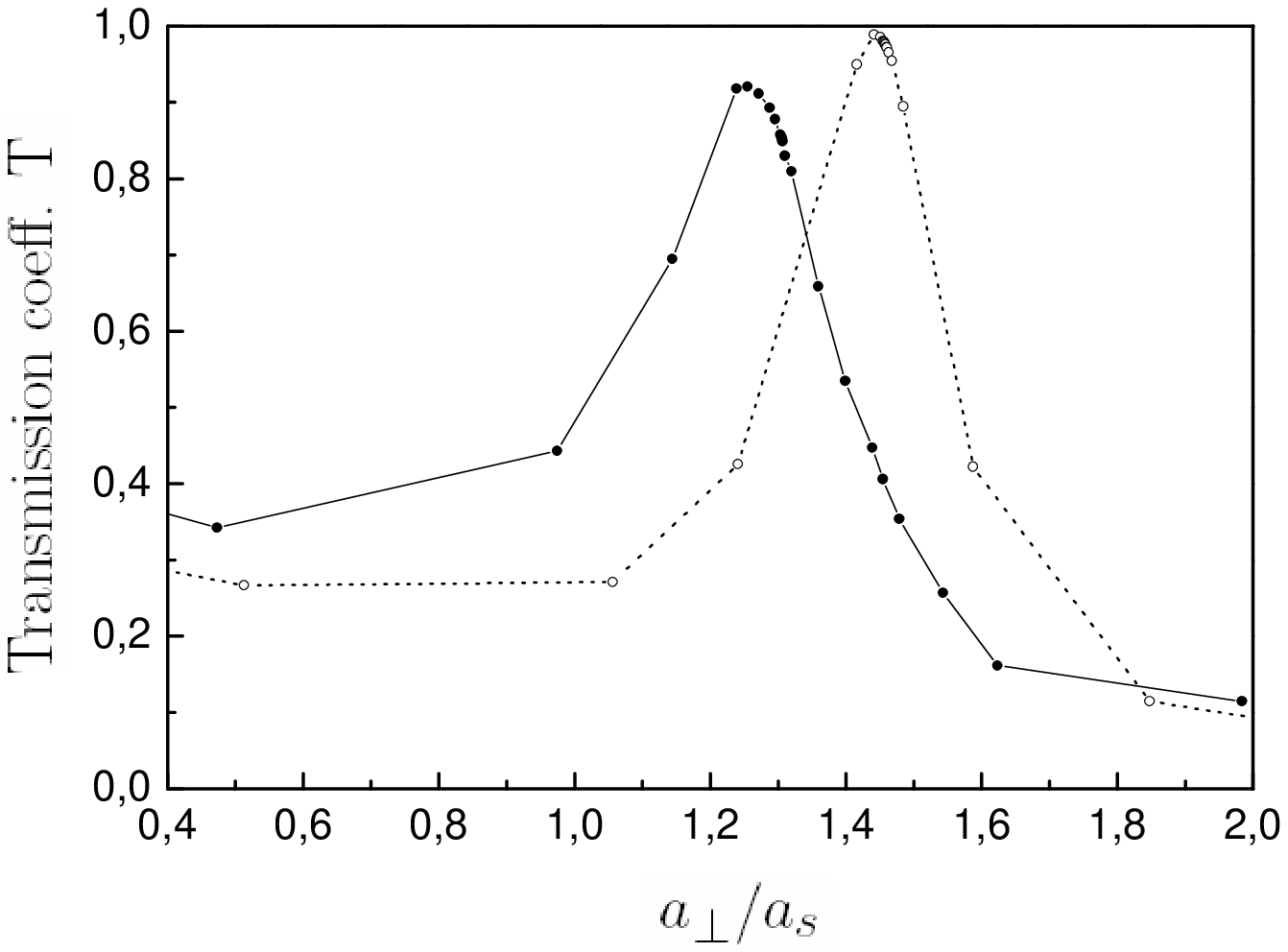}
\caption{
\label{fig3} The influence of the center of mass
coupling on the transmission T for $\omega_2=0.02$
and $\varepsilon=0.002$. Here $\omega_1= 1.35\omega_2$, $a_\perp/r_0=6.52$  (solid curve)
and $\omega_1=\omega_2$, $a_\perp/r_0=7.07$ (dotted curve). }
\end{figure}

This effect of the suppression of the quantum scattering
($T\approx 1$) for $-V_0\sim 8-9$ is encountered equally in the case of a
coupling with the center of mass $\omega_1\not=\omega_2$ (see Fig.~\ref{fig3}). The position
$a_{\perp}/ a_s\approx 1.45$ of the total transmission for
$\omega_1=\omega_2=0.02$ is only slightly shifted to
$a_{\perp}/a_s\approx 1.25 $ under the action of the coupling with
the center of mass for $\omega_1=1.35\omega_2$ and the maximum of
$T$ is slightly decreased. Note that in strong contrast to $T$,
the 3D cross section $\sigma$, e.g. for $V_0=-8.45$ and $E\sim \omega_1=\omega_2=0.02$,
can increase almost four-fold if
the $p$-wave contribution is added.

\begin{figure}[t]
\includegraphics[scale=0.70]{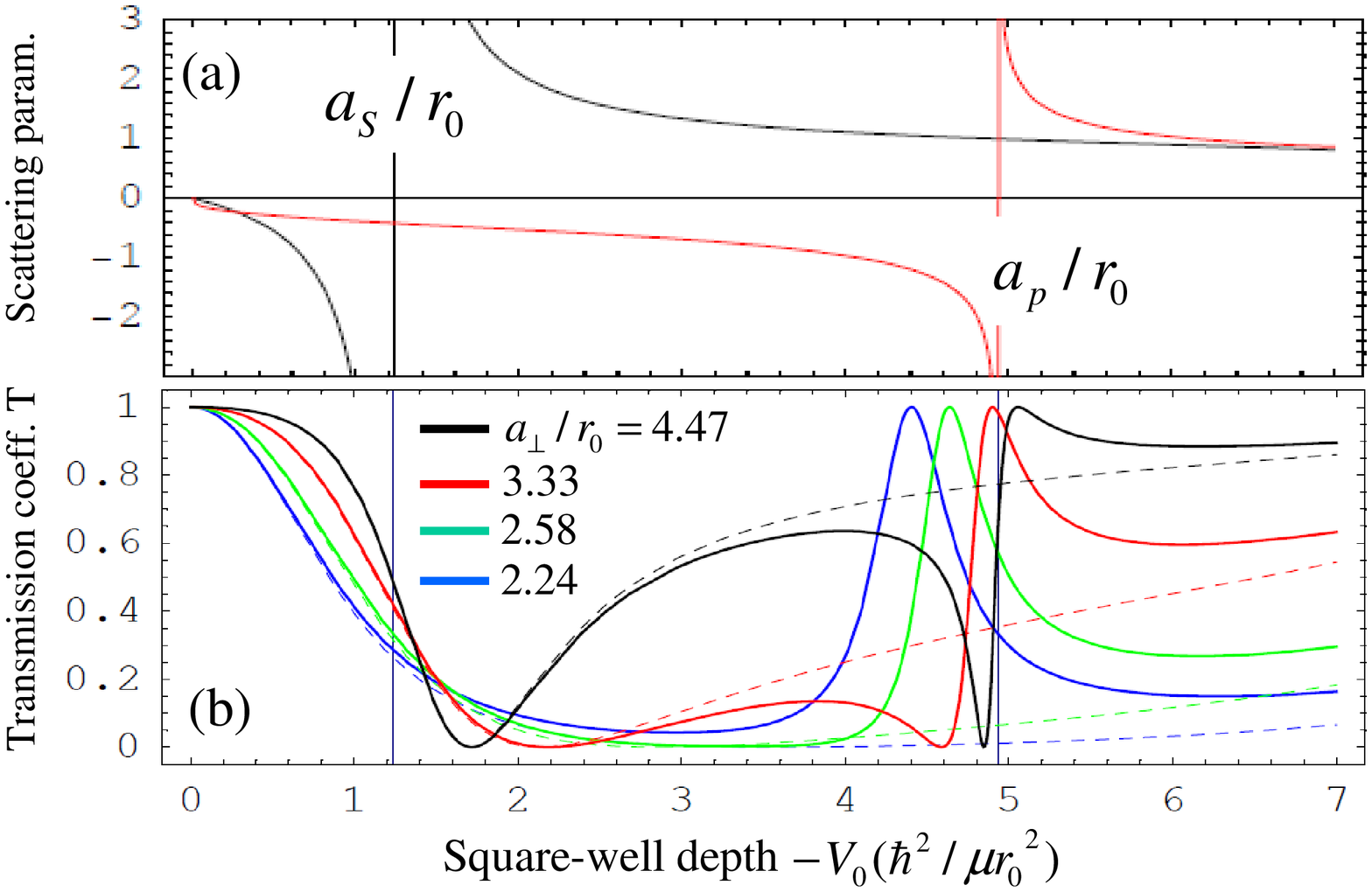}
\caption{ \label{fig4}
(color online) Scattering lengths and transmission for the square-well
potential (depth $V_0$ and radius $r_0$) for $\omega_1 = \omega_2$ and
$\varepsilon = 0.04$. ({\bf a}) $a_s = a_s(V_0)$ (black) and $a_p =
a_p(V_0)$ (red). ({\bf b}) $T = T(V_0)$ for several values of $a_\perp$
[see Fig.(1b)]. For comparison, $f_{0u}$ is omitted in the dashed
curves and only the s-wave CIR behaviour $T \approx 0$ due to large
$a_s$ appears.}
\end{figure}

The observed dramatic change of $T$ due to the combined action of $s-$ and
$p$-wave scattering is qualitatively confirmed in simpler
though solvable models, at least in the absence of the coupling to the
center of mass. Indeed, for $\omega_1=\omega_2$ and $k r_0=\sqrt{2\mu E} r_0\ll 1$
(i.e., low total energy), the condition equivalent to $f_{0u}\approx
-1$ can be obtained separately from the antisymmetric part of the
wave function as a divergence of a quasi-1D effective interaction
strength for
$a_\perp/a_p=-1.36$~\cite{granger2004a,girardeau2005a}. Together
with $a_\perp=1.46a_s$~\cite{olshanii1998a}, one finds again the
requirement of both large $a_s$ and $a_p$. Another more direct
determination of both $f_{0g,u}$ using the approximation developed in ~\cite{kim2005a}
predicts $a_\perp=2a_s=-2a_p$, namely, $f_{0g} = - (1 +
i\cot{\delta_g})^{-1}$ and $f_{0u} = - (1 +
i\cot{\delta_u})^{-1}$, where $\cot{\delta_g} = - [a_\perp/a_s - (
C^2 - a_\perp^2k_0^2 )^{1/2}] a_\perp k_0/2$ while $\cot{\delta_u}
= -[ a_\perp^3/V_p + ( C^2 - a_\perp^2k_0^2 )^{3/2} ] / (6a_\perp
k_0)$ is a straightforward improvement to Eq.(30b) of
Ref.~\cite{kim2005a} and $C=2$. Using then $a_s(V_0)$ and
$a_p(V_0)$ of a square-well of depth $V_0$ and radius $r_0$ in
the scattering amplitudes $f_{0g,u}$ we obtain $T$ from Eq.(\ref{t})
which is shown in Fig.~\ref{fig4}.
The resulting behavior is qualitatively
similar to Fig.~\ref{fig2}, confirming the role of both large $a_s$ and $a_p$ , on the
order of $a_\perp$.

For an experiment, the main condition for $T\approx 1$ is
$f_{0g,u}\approx -1$. If the partial waves $l \ge 2$ are
negligible and $k_0r_0 \ll kr_0 \ll 1$, this results in  the fact
that $a_s$ and $-a_p$ are of the order of $a_\perp$. More precise
values of these ratios~\cite{olshanii1998a,granger2004a} may
depend on details of the confining potential, coupling to the CM,
$k_0$, and should be computed for a given experimental setup.
Current laser traps for e.g. $^{40}$K can well reach the ranges
$a_s\sim a_\perp< 60$nm~\cite{moritz2005a} or $-a_p\sim
a_\perp<60$nm~\cite{guenter2005a}, using, however, {\em separate}
3D Feshbach resonances in the $s$- and
$p$-waves~\cite{ticknor2004a}. In a recent paper~\cite{krems2006a}
a mechanism for the {\em simultaneous} creation of s- and p-wave
Feshbach resonances is established: Applying laboratory
dc-electric fields introduces the herefore necessary mixing of s-
and p- waves. Thus, a reduction of quasi-1D scattering relative to
3D, as revealed e.g. in trap losses, should be observable with
tighter traps by further exploring Feshbach resonances. Another
example is the potential $V=-(Z_ie^2/\kappa r)e^{-r/r_0}$ of an
ionized impurity of valence $Z_i$ in a semiconductor quantum wire
with dielectric constant $\kappa$. Restoring the length unit
$r_0$, the relation to $V_0$ (in units of $\hbar^2/\mu r_0^2$) is
thus $r_0=-\kappa\hbar^2V_0/Z_i\mu e^2$. For $V_0=-8.45$,  the
required screening length $r_0$ can range from $5.4\,$nm (heavy
holes in Si and $Z_i=2$) up to $86.8\,$nm (electrons in GaAs and
$Z_i=1$). Geometrically, this range and $a_\perp\gg r_0$ are well
within current scaling technologies~\cite{cui2003a}.

In conclusion, due to quantum interference of different partial wave amplitudes, the scattering
in a quasi-1D geometry can be strongly suppressed, although the interaction or scattering in 3D is
strong.

J.I.K. acknowledges the Alexander von Humboldt Foundation for a scholarship.
Financial support by the Landesstiftung Baden-W\"urttemberg in the framework of
the project 'Mesoscopics and atom optics of small ensembles of ultracold atoms'
is gratefully acknowledged by P.S. and V.S.M.

\end{document}